# Tunneling electroresistance effect in ferroelectric tunnel junctions at the nanoscale

A. Gruverman,[1] D. Wu,[2] H. Lu,[1] Y. Wang,[1] H. W. Jang,[3] C.M. Folkman,[3] M. Ye. Zhuravlev,[1,4] D. Felker,[3] M. Rzchowski,[3] C.-B. Eom[3] and E. Y. Tsymbal[1]

[1]*University of Nebraska, Lincoln, NE 68588*

[2]*North Carolina State University, Raleigh, NC 27695*

[3]*University of Wisconsin-Madison, WI 53706*

[4]*Kurnakov Institute for General and Inorganic Chemistry, RAS, Moscow, Russia*

Abstract

Stable and switchable polarization of ferroelectric materials opens a possibility to electrically control their functional behavior. A particularly promising approach is to employ ferroelectric tunnel junctions where the polarization reversal in a ferroelectric barrier changes the tunneling current across the junction. Here, we demonstrate the reproducible tunneling electroresistance effect using a combination of Piezoresponse Force Microscopy (PFM) and Conducting Atomic Force Microscopy (C-AFM) techniques on nanometer-thick epitaxial $BaTiO_3$ single crystal thin films on $SrRuO_3$ bottom electrodes. Correlation between ferroelectric and electronic transport properties is established by the direct nanoscale visualization and control of polarization and tunneling current in $BaTiO_3$ films. The obtained results show a change in resistance by about two orders of magnitude upon polarization reversal on a lateral scale of 20 nm at room temperature. These results are promising for employing ferroelectric tunnel junctions in non-volatile memory and logic devices, not involving charge as a state variable.

Ferroelectrics comprise a class of polar materials where a spontaneous electric polarization can be switched by an applied electric field, which opens a possibility of electrical control of their functional properties. One of the particularly promising aspects is a switching of resistance in ferroelectric tunnel junctions (FTJ) upon polarization reversal [1]. A mechanism of electrically-induced resistance switching of FTJs, known as the tunneling electroresistance (TER) effect, is fundamentally different from those observed in a number of metal oxides, including titanates, manganites and zirconates. In these materials, depending on the specific system, the underlying mechanisms are quite diverse [2] and can include, among others, electromigration, crystalline defects [3], Schottky barrier variations [4] and electrically-induced filament formation [5]. A common feature of these mechanisms is that they are related to nano- or atomic scale inhomogeneities and do not involve ferroelectric polarization reversal [6, 7]. In the FTJs, where a ferroelectric film serves as a barrier between two metal electrodes, the resistance switching is realized via a pure electronic mechanism. If the ferroelectric film is sufficiently thin (of the order of a few nanometers), conduction electrons can quantum-mechanically tunnel through the ferroelectric barrier. By flipping the polarization of the ferroelectric barrier it is possible to change an internal electronic potential profile and, hence, alter the transmission probability and produce the TER effect [8,9,10]. Previous studies have observed signatures of phenomena similar to TER, however, no correlation between the polarization orientation and the tunneling conductance was established [11,12].

Realization of ferroelectric tunnel junctions relies on thermodynamic stability and switching in ultra-thin ferroelectric films. The problem of stability has been studied both theoretically and experimentally [13,14]. The equilibrium polarization state strongly depends on extrinsic factors such as film-substrate stress and screening conditions. Compressive strain due to the lattice mismatch was shown to lead to a significant increase of the spontaneous polarization and phase transition temperature in $BaTiO_3$ films [15]. Stabilization of the polar phase by reducing the depolarizing field energy has been achieved via formation of nanometer-period antiparallel 180º stripe domains [16] or by effective control of chemical environment that allows efficient screening of the polarization charges [17]. These results open a way for experimental implementation of the FTJ structures.

Structurally and electrically perfect epitaxial ferroelectric heterostructures with a thickness of just several unit cells are a prerequisite for implementation of FTJs. To achieve the atomically abrupt and well-defined interfaces, epitaxial $BaTiO_3$ films have been fabricated by atomic layer controlled growth with *in-situ* monitoring using high pressure reflection high energy electron diffraction (RHEED) [15, 18]. To probe the nanoscale switching behavior and transport we used epitaxial ferroelectric films of $BaTiO_3$ with a thickness in the range from 2.4 nm to 9.6 nm grown on atomically smooth $TiO_2$-terminated (001) $SrTiO_3$ substrates with single crystalline $SrRuO_3$ electrodes. Figure 1a shows RHEED intensity oscillation and RHEED patterns indicating a layer-by-layer growth and epitaxial structure of the $BaTiO_3$ films. The AFM image of completed $BaTiO_3$/$SrRuO_3$ heterostructure in Fig. 1b clearly indicates atomically-flat terraces with single unit-cell high (~4 Å) steps. Cross-sectional TEM images (not shown) demonstrate that the $BaTiO_3$ and $SrRuO_3$ layers are fully commensurate with the $SrTiO_3$ substrates.

We investigate the polarization-dependent TER effect by monitoring and controlling the electric polarization and correlating local ferroelectric switching with the local conductivity. Our approach is to employ a combination of scanning probe microscopy techniques for nanoscale polarization detection (Piezoresponse Force Microscopy, or PFM) [19, 20] and spatially-resolved local conductance measurements (Conducting Atomic Force Microscopy, or C-AFM) [21, 22]. The experimental geometry shown in Fig. 2a involves a conductive probing tip in contact with a bare BaTiO$_3$ surface (in the area of about 20 nm in diameter) connected to a power supply for local piezoresponse and current measurements.

The measurements have been performed using a commercial atomic force microscope AFM (Asylum MFP-3D). Resonant-enhanced PFM mode was used to obtain images of polarization patterns in BaTiO$_3$ films. Typical frequency range for an ac voltage was 350-400 kHz with an amplitude of 0.5 V (peak-to-peak). Conductive Pt-Ti-coated silicon cantilevers (NSC14, Mikromasch) were used for PFM imaging and polarization switching studies. Local piezoelectric hysteresis loops were measured in fixed locations on the film surface as a function of a dc switching bias superimposed on ac modulation bias.

Nanoscale electrical transport measurements have been performed in the C-AFM mode using a voltage source measure unit (Keithley 237) allowing reliable current detection in the range from $10^{-11}$ to $10^{-3}$ A. Local current-voltage characteristics were measured by positioning a conductive diamond tip (CDT-NCHR, Nanoworld AG) at a selected point on the film surface and by sweeping a dc voltage. Current maps were obtained simultaneously with the topographic data by scanning the surface with the tip held under a constant dc bias (not higher than 30% of the coercive voltage) and measuring current at each pixel point of the scanned area. Typical imaging scan rate was 0.8 Hz.

Figure 2b shows a local PFM response of a 4.8nm-thick BaTiO$_3$ film that reveals a hysteretic behavior typical for polarization switching. Note that the hysteresis loop is characterized by a shift toward a positive voltage indicating a built-in electric field of about $10^6$ V/cm. The direction of the field is upward (i.e. from the substrate to the tip) which is consistent with a single-domain state of as-grown BaTiO$_3$ films exhibiting the upward polarization orientation.

To demonstrate the TER effect, first, we change the upward-polarized single-domain state of the as-grown BaTiO$_3$ film to a bidomain-patterned state. For this purpose, polarization in a 2.5x2.5 µm$^2$ square region of the film is switched downward by scanning the film surface with a tip biased with a voltage $V_{tip} = +3.0\,V$ that exceeds the coercive voltage for the film. Then, polarization within an area of 0.5x0.5 µm$^2$ square area in the center is switched back upward by applying a bias $V_{tip} = -3.0\,V$ to the tip. Figure 2c shows the resulting polarization pattern as measured by PFM. Antiparallel domains written this way demonstrate almost no relaxation 3 days after the switching, which suggests a highly stable and robust polarization.

The resistive switching behavior of the BaTiO$_3$ films is examined using a C-AFM technique. A current map is acquired by scanning the polarization-patterned area with a dc bias of 0.3 V. Figure 2d shows the resulting current pattern within the poled area where variations in contrast correspond to different conductivity. The striking feature evident from the comparison of Figs. 2c and 2d is a perfect correspondence of the spatial variations in local conductivity measured by C-AFM to the polarization domain pattern measured by PFM. The area with downward polarization shows significantly higher conductivity than the area with upward polarization with the resistance ratio between the two polarization states of about 7 and 83 for the 2.4 nm and 4.8 nm thick BaTiO$_3$ films, respectively. By performing several sequential PFM/current imaging series we find that a small dc bias used for conductance measurements does not perturb the ferroelectric state thus allowing multiple nondestructive polarization readouts.

To distinguish the ferroelectric-induced mechanism of resistive switching from other mechanisms [2], we have examined correlation between the onset of polarization reversal and a change in electrical conductance in the same locations. The BaTiO$_3$ surface was scanned with the tip under an incrementally increasing dc bias with simultaneous PFM detection of polarization. The gradual change in the contrast under a tip bias close to the coercive voltage seen in Fig. 3a is an indication of polarization reversal. Subsequently, the current map was acquired in the same area by C-AFM. As seen from Fig. 3b, the transition from high to low resistance states occurred exactly at the same bias where polarization undergoes reversal from the upward to downward direction. Resistance stays constant in the fully switched regions irrespective of what voltage is used for poling. This result is an unambiguous demonstration of the ferroelectric nature of the resistive switching.

Spatially-resolved imaging of polarization and tunneling current are complemented by local spectroscopic measurements which provide a quantitative insight into resistive switching [23]. I-V characteristics recorded for oppositely poled regions (Fig. 4) show a non-linear behavior typical for tunneling conductance in tunnel junctions and a drastic change in resistance. In the case of the 4.8-nm thick BaTiO$_3$ film the resistance changes by almost two orders of magnitude upon polarization reversal. These remarkable results clearly illustrate the giant TER effect in FTJs predicted earlier [8].

It has been demonstrated theoretically [8, 10] that a change in electric polarization direction, and hence a change in the relative Ti displacements in the BaTiO$_3$ cell, can affect the density of states at the electrode-barrier interfaces, as well as the interface dipoles and complex band structure of the barrier, primarily due to changing interfacial atomic positions and electronic structure. In the case that the interfaces differ, the changes on polarization reversal are also likely to differ, leading to a dependence of tunneling current on polarization direction. To interpret out results, we assume direct tunneling through the barrier, with interfacial effects and their change with polarization reversal by potential steps at the interfaces that depend on the polarization direction.

To model the observed TER effect we employ a simple model, involving a tunneling current through a trapezoidal potential barrier [24] whose profile depends on polarization orientation. The barrier has width $d$ and potential steps at the interfaces of $\phi_1$ and $\phi_2$, so that the potential energy across the barrier in the presence of bias voltage $V$ varies as $\phi(x,V) = \phi_1 + eV/2 + x(\phi_2 - eV - \phi_1)/d$. Using the WKB approximation and assuming that the applied voltage is not too large, so that $eV/2 < \phi_{1,2}$ and the barrier width is not too small, so that $d\sqrt{\frac{2m}{\hbar^2}\phi_{1,2}} \gg 1$, we obtain an analytic expression for the current density:

$$J \cong C \frac{\exp\left\{\alpha(V)\left[\left(\phi_2 - \frac{eV}{2}\right)^{3/2} - \left(\phi_1 + \frac{eV}{2}\right)^{3/2}\right]\right\}}{\alpha^2(V)\left[\left(\phi_2 - \frac{eV}{2}\right)^{1/2} - \left(\phi_1 + \frac{eV}{2}\right)^{1/2}\right]^2} \sinh\left\{\frac{3}{2}\alpha(V)\left[\left(\phi_2 - \frac{eV}{2}\right)^{1/2} - \left(\phi_1 + \frac{eV}{2}\right)^{1/2}\right]\frac{eV}{2}\right\} \quad (1)$$

where $C = -\frac{4em}{9\pi^2\hbar^3}$ and $\alpha(V) \equiv \frac{4d\sqrt{2m}}{3\hbar(\phi_1 + eV - \phi_2)}$.

Figure 4 shows a fit of the experimental I-V characteristics for the 4.8nm thick BaTiO$_3$ film, using Eq. (1) and the following parameters: $\phi_1 = 0.24 eV$ and $\phi_2 = 1.52 eV$ for polarization pointing up, $\phi_1 = 0.48 eV$ and $\phi_2 = 0.96 eV$ for polarization pointing down, $d$ = 4.8 nm and $m$ = $m_0$. The respective change in the potential energy profile is shown schematically in the inset of Fig. 4. According to these results, the reversal of polarization changes the potential energy difference $\phi_1 - \phi_2$ across the BaTiO$_3$ barrier from 1.28 $eV$ to 0.48 $eV$. This change of 0.8 eV is due to the change in the electrostatic potential associated with ferroelectric polarization reversal and associated reorientation of the depolarizing field. We note that $\phi_2 - \phi_1$ is higher for polarization pointing up than for polarization pointing down indicating that "holes" dominate in the tunneling conductance. This behavior is expected if the decay constant of the evanescent state in the tunneling barrier is increasing with energy, which is the case for the $\Delta_1$ band in BaTiO$_3$ [25].

The fitting predicts a potential drop of 1.28V acting in the initial polarization state of the as-grown sample. We interpret this result as a consequence of the screening charge accumulation on the film surface, which is shown to have a profound effect in stabilizing the polarization in ultrathin films [17]. The screening charge most likely is a bound charge accumulated within a thin (a few Angstroms) surface dielectric layer formed after the ferroelectric film has been exposed to air. This dielectric layer is expected to have a higher band gap and consequently it produces a higher potential barrier than BaTiO$_3$ which within our trapezoidal barrier model leads to the potential slope in the as-grown state. Assuming that the bound screening charges create an internal electric field that almost fully compensates the depolarizing field in the barrier in the as-grown state, we conclude from our fitting that the intrinsic electric field produces an electrostatic potential drop of 0.8V/2=0.4V across the BaTiO$_3$ layer. The presence of this field is consistent with our experimental results which indicate a shift in the hysteresis curve by about the same value in Fig. 2b.

In the Ohmic transport regime typical for a small bias voltage, Eq. (1) is reduced to

$$J \approx \frac{e^2}{2h} \frac{1}{2\pi d} \frac{\sqrt{2m}}{\hbar} \left( \phi_2^{1/2} + \phi_1^{1/2} \right) \exp\left[ -\frac{4\sqrt{2m}}{3\hbar} \frac{\phi_2^{3/2} - \phi_1^{3/2}}{\phi_2 - \phi_1} d \right] V . \qquad (2)$$

If the difference between potential steps $\Delta\phi \equiv \phi_2 - \phi_1$ is not too large compared to the average potential $\bar{\phi} \equiv (\phi_2 + \phi_1)/2$, so that $\Delta\phi/2 < \bar{\phi}$, this formula can be approximated by

$$J \approx \frac{e^2}{h} \frac{1}{2\pi d} \frac{\sqrt{2m\bar{\phi}}}{\hbar} \exp\left[-2\frac{\sqrt{2m\bar{\phi}}}{\hbar} d\right] V, \tag{3}$$

which is consistent with Simmons's formula at low bias voltage [26]. According to Eq. (3), the transport properties are entirely determined by the average potential barrier height and we can discuss the TER effect in terms of the change of $\bar{\phi}$ with polarization reversal. We assume that the barrier height changes by $\delta\phi$ from $\bar{\phi} - \delta\phi/2$ (high conductance state, current density $J_>$) to $\bar{\phi} + \delta\phi/2$ (low conductance state, current density $J_<$) and define the TER ratio by $TER \equiv (J_> - J_<)/J_<$. Assuming that $\delta\phi$ is small and $J_> \gg J_<$ we arrive at

$$TER \equiv \frac{J_> - J_<}{J_<} \approx \exp\left[\frac{\sqrt{2m}}{\hbar} \frac{\delta\phi}{\sqrt{\bar{\phi}}} d\right]. \tag{4}$$

In case of the 4.8 nm thick $BaTiO_3$ film, we find that $\bar{\phi} = 0.80 \, eV$ and $\delta\phi = 0.16$ eV resulting in $TER \approx 80$ consistent with our fit and experiment.

It is noteworthy that Eq. (4) predicts a dramatic increase in TER with increasing barrier thickness $d$. This increase is expected to be stronger than exponential due to the dependence of $\bar{\phi}$ and $\delta\phi$ on $d$ (through the polarization-induced potential variation across the barrier), resulting in the increasing $\delta\phi/\sqrt{\bar{\phi}}$ with film thickness. This tendency is indeed observed experimentally. If we assume that both $\bar{\phi}$ and $\delta\phi$ are constants, according to Eq. (4), an increase in film thickness by a factor of two is expected to produce a TER that is a square of the value at the smaller thickness. In our case, an increase from $d = 2.4$ nm to $d = 4.8$ nm enhances TER from 7 to 83, i.e. by a factor of two larger than what is expected from a simple exponential dependence.

In summary, we have demonstrated polarization-dependent resistive switching behavior of ultrathin films of BaTiO$_3$ at room temperature, manifesting the tunneling electroresistance effect predicted earlier. Upon polarization reversal, tunneling resistance changes by almost two orders of magnitude on a lateral scale of about 20 nm. Polarization retention is not affected by conductance measurements thus allowing multiple nondestructive polarization readouts and opening a possibility for application as non-charge based logical switches in nonvolatile memory devices.

This work was supported by the National Science Foundation (Grant No. MRSEC DMR-0820521), the Nanoelectronics Research Initiative of Semiconductor Research Corporation, ECCS-0708759 and the Office of Naval Research through grants N00014-07-1-0215.

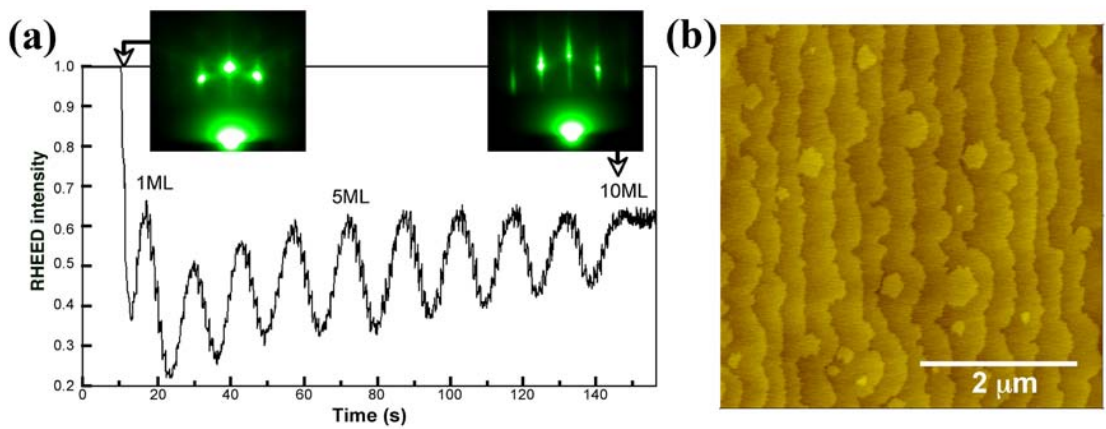

Fig. 1

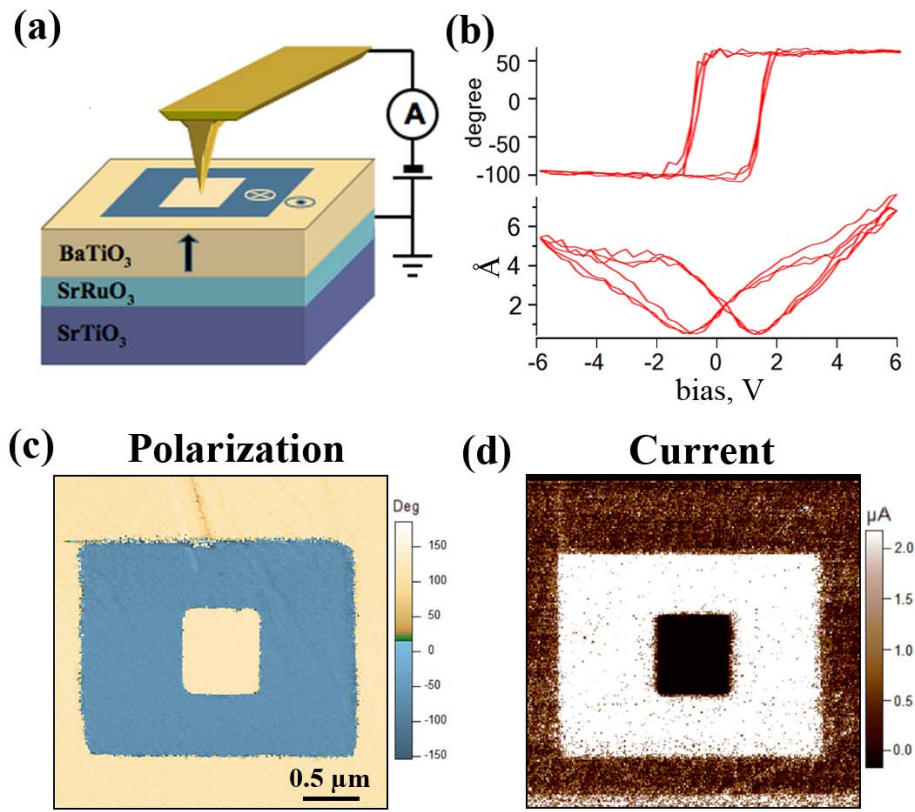

Fig. 2

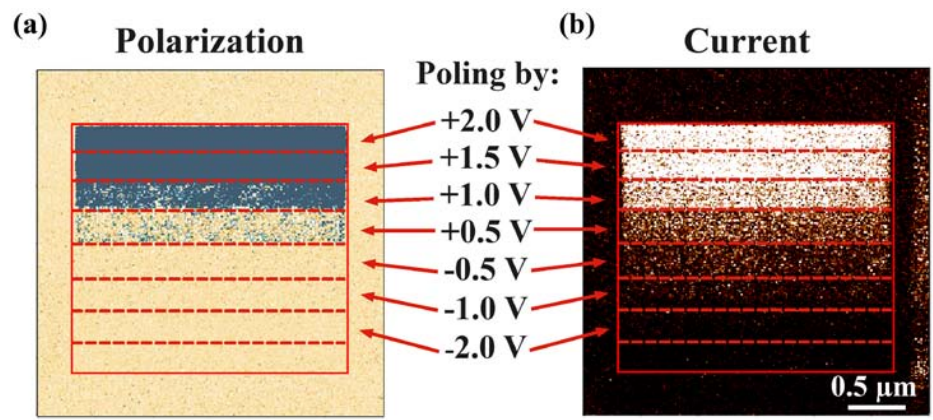

Fig. 3

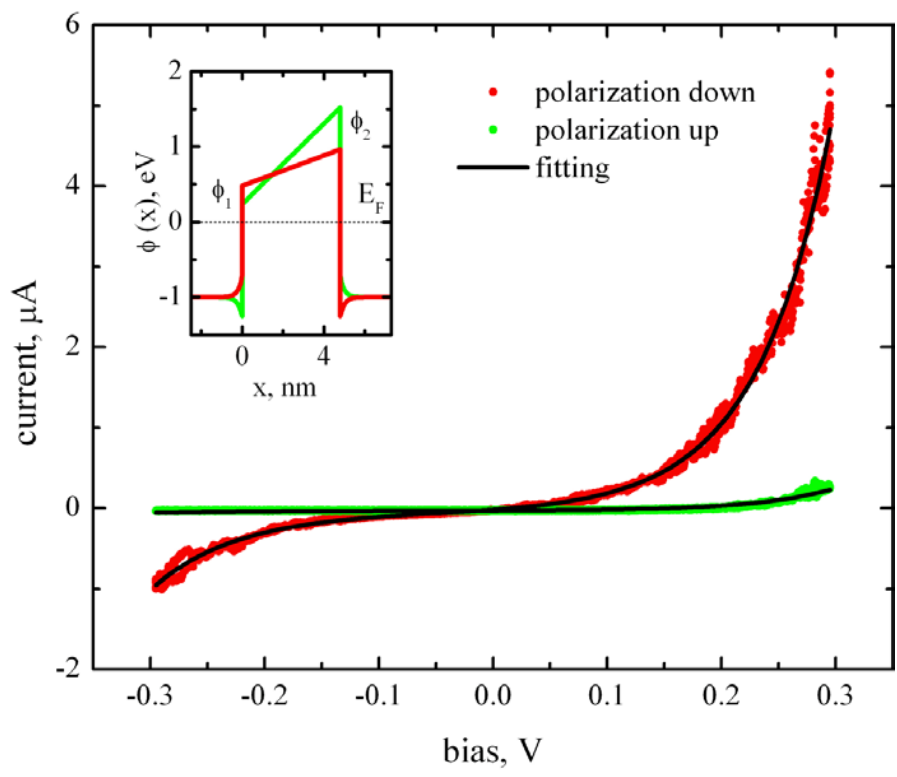

Fig. 4

**Figure Captions**

Figure 1. (a) RHEED intensity oscillation during the growth of BaTiO$_3$ ferroelectric layer on SrRuO$_3$ bottom electrode and RHEED patterns before and after deposition of the BaTiO$_3$ layer; (b) AFM image of the BaTiO$_3$ film on the SrRuO$_3$/SrTiO$_3$ substrate.

Figure 2. (a) Sketch illustrating the geometry of experiment for PFM/C-AFM studies of tunneling electroresistance effect in ultra-thin BaTiO$_3$ films. (b) Local PFM hysteresis loops measured in the 4.8-nm thick BaTiO$_3$ film (top - phase signal, bottom - amplitude signal). (c) PFM image of a polarization pattern produced by scanning with the tip under 3.0 V (blue region corresponds to polarization switched downward; yellow region represents upward polarization in the as-grown film). (d) Tunneling current map acquired in the same region as in (c). Bright contrast indicating higher current is observed in the regions with polarization oriented downward.

Figure 3. Spatially resolved correlation between the onset of polarization reversal (a) and a change in electrical conductance (b). A change in the polarization contrast in the red block from yellow to blue illustrates polarization reversal under an incrementally changing tip bias. Dashed red lines indicate where the bias is changing. The change in PFM contrast correlates with the transition from low current (dark contrast) to high current (bright contrast).

Figure 4. I-V curves for two opposite polarization directions in the 4.8-nm thick BaTiO$_3$ film measured by C-AFM. Solid lines - fitting of the experimental data by the WKB model. The inset shows schematically the potential energy profiles for two polarization orientations.